\documentclass[prl,aps,twocolumn,preprintnumbers,showpacs,superscriptaddress]{revtex4}
\usepackage[latin1]{inputenc}
\usepackage{amsmath}
\usepackage{amsfonts}
\usepackage{amssymb}
\usepackage{graphicx}

\newcommand{\Tr}{{\textrm{Tr}}}
\newcommand{\Sp}{{\textrm{Sp}}}

\begin{document}
\title{Generalized spectroscopy; coherence, superposition, and loss}
    \author{Johan \surname{{\AA}berg}}
    \email{J.Aberg@damtp.cam.ac.uk}
    \affiliation{Centre for Quantum Computation, Department of Applied
      Mathematics and Theoretical Physics, University of Cambridge, Wilberforce
      Road, Cambridge CB3 0WA,
    United Kingdom}
\author{Daniel K. L. \surname{Oi}}
\affiliation{Department of Physics, SUPA, University of Strathclyde, Glasgow
  G4 0NG, United Kingdom}

\begin{abstract}
  We analyze single particle coherence and interference in the presence of particle loss and
  derive an inequality that relates the preservation of coherence, the
  creation of superposition with the vacuum, and the degree of particle
  loss. We find that loss channels constructed using linear optics form a
  special subclass. We suggests a generalized spectroscopy where, in analogy
  with the absorption spectrum, we measure a ``coherence loss spectrum" 
  and a ``superposition creation spectrum". The theory is illustrated with examples.
\end{abstract}


\maketitle

Interference phenomena are versatile tools for studying quantum systems. The
coherence of a physical process, i.e., its ability to preserve the 
capacity for interference, can likewise provide valuable information on the
dynamics, which we suggest could be utilized for a spectroscopic approach. 
The coherence properties of quantum operations have been considered
in Refs.~\cite{Ann,Oi2003,Aaberg2003}, for an alternative approach see
\cite{Braun}, and used to define operational interferometric fidelity and
coherence measures \cite{OiAb}. However, these investigations assume no
particle loss, limiting their applicability to mainly idealized situations. In
this Letter, we consider the effect of particle loss on coherence and suggest
procedures to measure the quantum properties of loss processes.  The relation
between particle loss and coherence has been considered both theoretically and
experimentally in the context of neutron interferometry \cite{neutronexp}
using complex phase shifts, and also in the context of geometric phases \cite{GeomPh} 
using non-Hermitian Hamiltonians.  Complex phase shifts and
non-Hermitian Hamiltonians \cite{Wong} are useful phenomenological
approaches to particle loss, and the latter can be derived as an approximation describing
no-jump trajectories in the quantum jump approach~\cite{qjump}. However, within
the standard framework of quantum mechanics, we model loss using
quantum channels (trace preserving completely positive maps)~\cite{Kraus} on
second quantized systems to allow for varying particle number.

Briefly reviewing the ideal case of no particle loss, 
consider a particle with
an internal degree of freedom (e.g. spin or polarization) traversing a
Mach-Zehnder interferometer with two paths A and B. In path A we insert the
material or device we wish to probe. In path B we apply a variable unitary
operator $U$ and a phase shift $e^{i\chi}$. After re-interfering the two
paths, the probability to find the particle in path $A$ is $p_{A}= \frac{1}{2} +
\frac{1}{2}|F(\rho,U)|\cos(\arg F(\rho,U)-\chi)$, where $F(\rho,U)
=\Tr(U^{\dagger}V\rho)$, and where $\rho$ is the initial internal state of the
particle, assuming no particle loss. The operator $V$, which we refer to as
the coherence operator \cite{Oi2003}, is not uniquely determined by the
internal state evolution channel $\Phi$ of the material 
\cite{Ann,Oi2003,Aaberg2003}, and thus provides
additional information. In a second quantized description, restricted to the 
vacuum and single particle subspace, the action of the device or
material can be described as $\widetilde{\Phi}_{A}\otimes \widetilde{I}_{B}$
where $\widetilde{I}_{B}$ denotes the identity channel on path $B$, and where
$\widetilde{\Phi}_{A}(\widetilde{\rho}) = \Phi(\widetilde{\rho}) +
V\widetilde{\rho}|0_{A}\rangle\langle 0_{A}| + |0_{A}\rangle\langle
0_{A}|\widetilde{\rho} V^{\dagger} + |0_{A}\rangle\langle
0_{A}|\widetilde{\rho}|0_{A}\rangle\langle 0_{A}|$ \cite{Ann,Aaberg2003}. Hence, 
the coherence operator $V$ simultaneously determines to what extent 
superposition between the vacuum and single particle states is preserved, 
and the capacity for interference.

The additional information in the coherence operator suggests a spectroscopic
procedure, recording the coherence as a function of the wavelength of the
probing particle, to obtain a ``coherence loss spectrum'', akin to an
absorption spectrum. (Not to be confused with coherence spectroscopy \cite{cohspe}.) 
Here, we generalize our previous approaches to include
loss in order to characterize this type of spectroscopy. Clearly, loss of the
interfering particles causes a reduction of interference, but how much?  We
find an expression that relates the interference capacity with particle loss,
but also with a third quantity that describes the creation of superposition
between the vacuum and single particle states.  To
achieve this we need to quantify superposition \cite{superpm}. Given two
projectors $P_{0}$ and $P^{\perp}$ onto orthogonal subspaces, the function
$A(\rho) = ||P^{\perp}\rho P_{0}||$ \cite{superpm}, where $||\cdot||$ is the standard 
operator norm, quantifies the superposition in a state $\rho$ with respect to the two subspaces. 
In the present case, where the vacuum state results in the one-dimensional projector
$P_{0} = |0\rangle\langle 0|$, we find $||P^{\perp}\rho P_{0}||=
||P^{\perp}\rho|0\rangle||$, where on the right hand side we have the ordinary
Hilbert space norm.

Let the channel $\widetilde{\Phi}$ be ``vacuum preserving'', i.e.,
$\widetilde{\Phi}(|0\rangle\langle 0|) = |0\rangle\langle 0|$.  Let
$|\psi^{\perp}\rangle$ be a normalized element in the orthogonal complement of
the vacuum state.
We define three functions that characterize the action of
the channel $\widetilde{\Phi}$.  The first, $\mathcal{L}(\psi^{\perp}) = \langle
0|\widetilde{\Phi}(|\psi^{\perp}\rangle\langle\psi^{\perp}|)|0\rangle$, tells
us to what extent the state $|\psi^{\perp}\rangle$ is mapped to the vacuum
state, i.e., the degree of loss. The next function, $\mathcal{P}(\psi^{\perp}) =
||P^{\perp}\Phi(|\psi^{\perp}\rangle\langle 0|)P_{0}||$, describes how well
$\widetilde{\Phi}$ preserves superposition between $|0\rangle$ and $|\psi^{\perp}\rangle$. 
Finally, $\mathcal{C}(\psi^{\perp}) =
||P^{\perp}\Phi(|\psi^{\perp}\rangle\langle\psi^{\perp}|)P_{0}||$ quantifies
how much superposition the operation creates from the input
$|\psi^{\perp}\rangle$.  If $\widetilde{\Phi}$ is vacuum preserving, the
following relation holds
\begin{equation}
\label{relation}
\mathcal{L}(\psi^{\perp})\mathcal{P}^{2}(\psi^{\perp}) + \mathcal{C}^{2}(\psi^{\perp})
\leq \mathcal{L}(\psi^{\perp})[1-\mathcal{L}(\psi^{\perp})].
\end{equation}
Moreover, if $\mathcal{L}(\psi^{\perp})=0$, then $\mathcal{C}(\psi^{\perp})=0$.
 
To prove Eq.~(\ref{relation}) we note that there always exists a Hilbert space
$\mathcal{H}_{a}$ and a unitary operator $\mathbb{U}$ on
$\widetilde{\mathcal{H}}\otimes\mathcal{H}_{a}$ such that
$\widetilde{\Phi}(\rho) = \Tr_{a}(\mathbb{U}\rho\otimes|a\rangle\langle
a|\mathbb{U}^{\dagger})$. By the requirement that $\widetilde{\Phi}$ should be vacuum
preserving it follows that there exists a normalized
$|a_{0}\rangle\in\mathcal{H}_{a}$, such that $\mathbb{U}|0, a\rangle = |0,
a_{0}\rangle$ (where $|x,y\rangle = |x\rangle|y\rangle$). 
We define $|f\rangle = \mathbb{U}|\psi^{\perp},a\rangle$, 
and note that $\langle 0,a_{0}|f\rangle =0$.
We can make the
following identifications: $\mathcal{L}(\psi^{\perp}) = ||\langle
0|f\rangle||^{2}$, $\mathcal{P}(\psi^{\perp}) = ||\langle
a_{0}|f\rangle||$, and $\mathcal{C}(\psi^{\perp}) =
||P^{\perp}_{0}\Tr_{a}(|f\rangle\langle f|)P_{0}||$, where we keep in mind  that $\langle
0|f\rangle  \in\mathcal{H}_{a}$ and $\langle
a_{0}|f\rangle\in\widetilde{\mathcal{H}}$. We let
$P^{\perp}_{a_{0}}$ denote the projector onto the orthogonal complement of
$|a_{0}\rangle$, and note that we can write
$\mathcal{L}(\psi^{\perp}) = \langle f|P_{0}\otimes
P_{a_{0}}^{\perp}|f\rangle$ and
$\mathcal{P}^{2}(\psi^{\perp}) = \langle f|P_{0}^{\perp}\otimes
P_{a_{0}}|f\rangle$. Finally, one can show that $\mathcal{C}^{2}(\psi^{\perp})\leq \langle
P_{0}^{\perp}\otimes P_{a_{0}}^{\perp}|f\rangle\langle f|P_{0}\otimes
P_{a_{0}}^{\perp}|f\rangle$.
 We can now prove Eq.~(\ref{relation}) by using the
identity $\langle f|P_{0}^{\perp}\otimes P_{a_{0}}^{\perp}|f\rangle + \langle
f|P_{0}^{\perp}\otimes P_{a_{0}}|f\rangle + \langle f|P_{0}\otimes
P_{a_{0}}|f\rangle =1$.  We can interpret Eq.~(\ref{relation}) as an exclusion
principle satisfied by vacuum preserving channels; for a given level of loss
the preservation of superposition and the creation of superposition are
competing, one takes its maximum only if the other is zero.

The relation in Eq.~(\ref{relation}) is valid for arbitrary vacuum preserving
operations, no matter how they act on the orthogonal complement of the vacuum
state (e.g., single particle states may be mapped to two-particle
states). However, a clear relation between $\mathcal{P}(\psi^{\perp})$ and
coherence, as measured with a single particle interferometer, requires the
assumption that single particle states are not mapped outside the
vacuum-single particle subspace, e.g., if $\widetilde{\Phi}$ has no gain. With
this additional assumption, we can further understand Eq.~(\ref{relation}) by
considering a particle without internal degree of freedom. In this case we can
explicitly construct the loss channels on the vacuum-single particle
states as
\begin{equation}
\label{endim}
\widetilde{\Phi}(\rho)  =   |0\rangle\langle 0|\rho_{00} + \sigma\rho_{11}
 + \gamma|0\rangle\langle 1|\rho_{01} + \gamma^{*}|1\rangle\langle 0|\rho_{10},
\end{equation}
where $\sigma$ is a density operator on $\Sp\{|0\rangle,|1\rangle\}$, 
and $\gamma$ is a complex number, such that $|\gamma|\leq 1$ and 
$\sigma_{00}|\gamma|^{2} + |\sigma_{01}|^{2} \leq \sigma_{00}(1-\sigma_{00})$. 
Hence, $\mathcal{L} = \sigma_{00}$, $\mathcal{P} = |\gamma|$, 
and $\mathcal{C} = |\sigma_{01}|$. 
One can show that the converse also holds; if $\sigma$ is a density operator
then $\widetilde{\Phi}$ defined by Eq.~(\ref{endim}) is a vacuum preserving
channel on the vacuum and single particle states.

How do we measure $\mathcal{L}(\psi^{\perp})$, $\mathcal{P}(\psi^{\perp})$,
and $\mathcal{C}(\psi^{\perp})$?  We first note that
$\mathcal{L}(\psi^{\perp})$ can be measured directly from the probability of
finding the vacuum in
$\widetilde{\Phi}(|\psi^{\perp}\rangle\langle\psi^{\perp}|)$. If
$\widetilde{\Phi}(|\psi^{\perp}\rangle\langle\psi^{\perp}|)$ stays within the
 vacuum and single particle subspace, then
$\mathcal{P}(\psi^{\perp})$, with $|\psi^{\perp}\rangle $ a single-particle
state, can be measured using a Mach-Zehnder setup, where the internal input
state of the particle is $|\psi^{\perp}\rangle$. We find that the final
probability to detect the particle in path $A$ is $p_{A}(\chi) = \frac{1}{2}
-\frac{1}{4}\mathcal{L}(\psi^{\perp}) +\frac{1}{2}|F|\cos(\arg F-\chi)$, where $F =\langle
\psi^{\perp}|U^{\dagger}\widetilde{\Phi}(|\psi^{\perp}\rangle\langle
0|)|0\rangle$. Thus, the maximal visibility
$\frac{1}{2}\mathcal{P}(\psi^{\perp})$ is obtained when
$U|\psi^{\perp}\rangle$ is parallel to
$P^{\perp}\widetilde{\Phi}(|\psi^{\perp}\rangle\langle 0|)|0\rangle$. It is difficult
to see how $\mathcal{C}(\psi^{\perp})$ could be measured using an
interferometric setup. However, there are other means to determine this
quantity, at least if the particle is a photon. There are
experimental techniques~\cite{Hessm} to determine the probability $p_{\chi} =
\langle \chi|\sigma |\chi\rangle$, where $|\chi\rangle = (|0\rangle +
e^{-i\chi}|1\rangle)/\sqrt{2}$ for $\chi\in [0,2\pi)$, and where $\sigma$ is a
density operator on the vacuum and single particle states. It is thus possible
to determine $\sigma_{01}$ by varying $\chi$. We can apply $\widetilde{\Phi}$
on a single particle state $|\psi^{\perp}\rangle$ and use the above technique to estimate
$p_{\chi}$, and thus determine $|\sigma_{01}|$. This approach can in
principle be extended to take into account an internal degree of freedom such that 
we can measure $\mathcal{C}(\psi^{\perp})$.

Let us examine two examples. First, consider a beam-splitter with transmissivity
$\cos^2\theta$,  coupling the mode of interest with an
ancillary mode initially in the vacuum state. The resulting channel
$\widetilde{\Phi}_{\textrm{bs}}$ on vacuum-single particle subspace of the mode 
of interest is as in
Eq.~(\ref{endim}) with $\sigma = \sin^{2}(\theta)|0\rangle\langle 0| +
\cos^{2}(\theta)|1\rangle\langle 1|$ and $\gamma = \cos(\theta)$. We
have no superposition creation, $\mathcal{C} \equiv |\sigma_{01}| =0$,  
and $|\gamma|$ takes its maximal value for a
given degree of loss.  As a second example we let the beam-splitter be either
totally transparent ($\theta = 0$) with probability $p$ or  totally reflective 
($\theta =\pi/2$) with probability $1-p$. On average we obtain a
channel $\widetilde{\Phi}_{\textrm{rand}}$ as in Eq.~(\ref{endim}), with
$\sigma = p|1\rangle\langle 1| + (1-p)|0\rangle\langle 0|$ and $\gamma = p$.
For comparison we let $p = \cos^{2}(\theta)$, yielding the same $\sigma$ 
as previously, but $\gamma =
\cos^{2}(\theta)$. Thus, for the same level of particle
loss, $\widetilde{\Phi}_{\textrm{rand}}$ preserves less coherence than
$\widetilde{\Phi}_{\textrm{bs}}$. 
In Ref.~\cite{neutronexp} (with a potentially confusing terminology mismatch 
with this Letter), ``deterministic" absorption using a beam chopper is equivalent 
to $\widetilde{\Phi}_{\textrm{rand}}$, and ``stochastic" absorption by the foil 
absorber we associate with $\widetilde{\Phi}_{bs}$. 

Beam splitters are examples of linear optics. Here we
consider the more general question of which vacuum preserving channels can be
obtained using linear optics only. We consider $K$  bosonic system modes 
$\mathcal{A}_{1},\ldots,\mathcal{A}_{K}$ (corresponding, e.g., to an internal 
degree of freedom of the particle) with
annihilation operators $\boldsymbol{a} = (a_{1},\ldots, a_{K})$, and $J$ ancillary
modes $\mathcal{B}_{1},\ldots,\mathcal{B}_{J}$ with annihilation operators
$\boldsymbol{b} = (b_{1},\ldots, b_{J})$. On the total system we assume linear
optics \cite{linopt}, which we describe with a unitary operator $\mathbb{U}$ such that
\begin{equation}
\label{Udef}
\mathbb{U}^{\dagger}\left[\begin{matrix}\boldsymbol{a}\\
\boldsymbol{b}
\end{matrix}\right]\mathbb{U} = \boldsymbol{S}\left[\begin{matrix}\boldsymbol{a}\\
\boldsymbol{b}
\end{matrix}\right],\quad \boldsymbol{S}=\left[\begin{matrix}\boldsymbol{S}^{(11)}
  &\boldsymbol{S}^{(12)}\\
  \boldsymbol{S}^{(21)} & \boldsymbol{S}^{(22)}
 \end{matrix}\right],
\end{equation}
where $\boldsymbol{S}$ is a unitary matrix.  We wish to find all vacuum
preserving channels $\widetilde{\Phi}(\rho) = \Tr_{e}(\mathbb{U}\rho\otimes\eta
\mathbb{U}^{\dagger})$, where $\eta$ is an arbitrary but fixed density operator on the
ancillary modes.  Let $\boldsymbol{S}^{(12)} =
\boldsymbol{V}\boldsymbol{D}\boldsymbol{W}^{\dagger}$ be a singular value
decomposition, i.e., $\boldsymbol{V}$ and $\boldsymbol{W}$ are unitary
matrices, and $\boldsymbol{D}$ is such that $\boldsymbol{D}_{ll'} =
d_{l}\delta_{ll'}$ for $1\leq l,l'\leq \min(K,J)$ and zero otherwise, and $d_{l}\geq 0$. 
We transform to new modes $\overline{\mathcal{A}}$ and
$\overline{\mathcal{B}}$, with annihilation operators
$\overline{\boldsymbol{a}} = \boldsymbol{V}^{\dagger}\boldsymbol{a}$ and
$\overline{\boldsymbol{b}} = \boldsymbol{W}^{\dagger}\boldsymbol{b}$,
respectively. We find that $\mathbb{U}^{\dagger}\overline{\boldsymbol{a}}\mathbb{U} =
\boldsymbol{V}^{\dagger}\boldsymbol{S}^{(11)}\boldsymbol{a} +
\boldsymbol{D}\overline{\boldsymbol{b}}$. 
(Note $\boldsymbol{a}$, not $\overline{\boldsymbol{a}}$, on the right hand side.)  
Since the channel $\widetilde{\Phi}$ is supposed to be vacuum preserving it follows that
$\Tr[\overline{n}_{k}\widetilde{\Phi}(|0\rangle\langle 0|)] =0$, where
$\overline{n}_{k} = \overline{a}_{k}^{\dagger}\overline{a}_{k}$.  Without loss
of generality we may assume  $\eta =
|\psi\rangle\langle\psi|$, yielding $ \langle
0,\psi|\mathbb{U}^{\dagger}\overline{n}_{k}\mathbb{U}|0,\psi\rangle =
d_{k}||\overline{b}_{k}|\psi\rangle||^{2}$, and hence $d_{k}\overline{b}_{k}|\psi\rangle =0$.  Now we
consider the action of the channel $\widetilde{\Phi}$ on the vacuum and single
particle states and use the above results to find:
\begin{eqnarray}
\label{linoptkan}
\widetilde{\Phi}(\rho) & = &  |0\rangle\langle 0|\Tr[(\hat{1}-S^{\dagger}S)\rho]
+S\rho S^{\dagger}\nonumber\\
& & + S\rho|0\rangle\langle 0| + |0\rangle\langle 0| \rho S^{\dagger},
\end{eqnarray}
 where $S = \sum_{l,l'=1}^{J} |\mathcal{A}_{l}\rangle
\boldsymbol{S}^{(11)}_{ll'}\langle \mathcal{A}_{l'}|$, and where 
$|\mathcal{A}_{l}\rangle$ denotes the single particle excitation in
mode $\mathcal{A}_{l}$. A necessary and sufficient condition for the channel 
in Eq.~(\ref{linoptkan}) to be obtainable via linear optics is 
that  $SS^{\dagger}\leq P_{1}$ and $S^{\dagger}S\leq P_{1}$, 
with $P_{1}$ the projector onto the single-particle subspace. 
One can see that $\mathcal{P}^{2}(\psi^{\perp}) =
1- \mathcal{L}(\psi^{\perp})$ and $\mathcal{C}(\psi^{\perp}) = 0$. 
Hence, the coherence preservation is maximal relative to the degree of particle loss, and
there is no superposition creation.  Note that excessive coherence loss is
obtained if we form convex combinations of linear optics channels, i.e. by
random selection of ``pure" linear optics channels. We can also conclude that, 
even at the level of vacuum and single particle states, 
the linear optics channels form only a subset of all possible channels. 

\begin{figure}
\includegraphics[width = 0.45\textwidth]{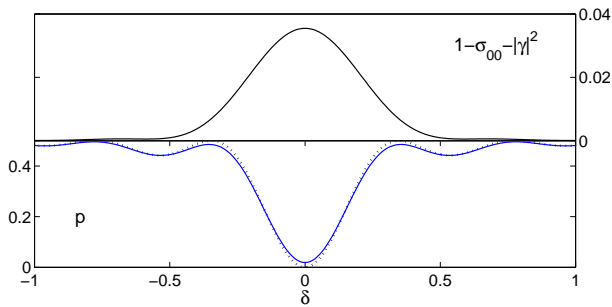}
\caption{\label{fig:absorb} (Color online) A photon interacting
  with a dephasing two-level atom via the Jaynes-Cummings model. The solid
  (dotted) line in the lower panel depicts the probability that a photon is
  found in the field, as a function of the detuning $\delta = \omega-\omega_{a}$, 
  at time $t = \pi/(2g)$ with (without) dephasing. The upper panel
  shows the excess coherence loss $1-\sigma_{00}-|\gamma|^{2}$ as a function
  of detuning $\delta$ in the case with dephasing. Neither the pure
  Jaynes-Cummings model, nor a simple relaxation model for the atom, 
  gives any excess coherence loss.}
\end{figure}

Now we turn to the question of coherence loss spectra, and as a model system
 we consider a photonic mode and a
two-level atom interacting via a Jaynes-Cummings Hamiltonian \cite{JC}. 
On the relevant subspace  
we can write the effective total Hamiltonian as
\begin{equation}
\label{JC}
H = \frac{\omega}{2}\sigma_{z}\otimes \hat{1}_{a}
+\frac{\omega_{a}}{2}\hat{1}\otimes\sigma_{z}^a + g(|01\rangle\langle 10|
+|10\rangle\langle 01|),
\end{equation}
where $\omega$ is the photon energy, $\omega_{a}$ the excitation
energy of the atom, and $g$ the coupling constant, all in units of some 
suitable reference energy $\mathcal{E}$, where we assume $\hbar = 1$. 
If the atom initially is in the ground state then the resulting
 channel on the vacuum and single particle subspace of the field is such that 
$|\sigma_{01}|=0$ and $|\gamma|^{2} =
1-\sigma_{00}$. Hence, there is neither generation of superposition nor
excessive coherence loss. We now assume the atom is affected by an 
environment, modeled with the master equation $\frac{d}{dt}\rho = -i[H,\rho]
+ q\mathcal{Q}(\rho)$, where $q\geq 0$, and where the 
time-parameter $t$ is in units of $\mathcal{E}^{-1}$. 
We consider two cases
\begin{eqnarray}
\label{Qdef}
\mathcal{Q}_{r}(\rho) & = &
|0\rangle_a\langle 1|\rho|1\rangle_a\langle 0|-\frac{1}{2}|1\rangle_a\langle 1|\rho
-\frac{1}{2}|1\rangle_a\langle 1|\rho,\nonumber\\
\mathcal{Q}_{d}(\rho) & = &
-\frac{1}{4}[\sigma_{z}^a,[\sigma_{z}^a,\rho]],
\end{eqnarray}
where $\mathcal{Q}_{r}$ gives relaxation to the ground state of the atom, and
$\mathcal{Q}_{d}$ gives dephasing in its eigenbasis.  For both
these cases $\sigma_{01}=0$. Hence, we can take $1-\sigma_{00}-|\gamma|^{2}$
as a measure of excess coherence loss. One can show that for the relaxation
model there is no excessive coherence loss, but there is for dephasing. 
We let $g = 0.1$, $\omega_{a} = 1$, and $q = 0.01$.
The upper panel of Fig.~\ref{fig:absorb} shows the excessive
coherence loss $1-\sigma_{00}-|\gamma|^{2}$ as a function of the detuning
$\delta = \omega-\omega_{a}$, at $t=\pi/(2g)$. In the lower panel the solid (dotted) line depicts
the probability $p$ to find a photon in the field as a function of the detuning
delta $\delta$, at the same $t$ with (without) dephasing. The excessive
coherence loss spectrum distinguishes dephasing from
the pure Jaynes-Cummings
interaction and relaxation, and hence may discriminate atom-environment couplings.

So far, all examples have resulted in $\mathcal{C} = 0$. Here we show how 
$\mathcal{C}\neq 0$ can occur.  If we can create the superposition
$|\chi\rangle = (|0\rangle + |1\rangle)/\sqrt{2}$, then the channel
$\widetilde{\Phi}(\rho) = |0\rangle \langle 0|\rho|0\rangle\langle 0| +
|\chi\rangle\langle\chi|\langle 1|\rho|1\rangle$ can be generated.  We
measure the particle number in the mode: if we measure vacuum we do nothing,
otherwise we prepare $|\chi\rangle$. The measurement destroys any initial
superposition, but as Eq.~(\ref{relation}) shows, this is the price we have to
pay for maximal superposition creation. Whether it is possible to generate the
superposition $|\chi\rangle$ depends on the nature of the particle. If the 
particle is a photon then the superposition can be generated using, e.g., 
photon blockade~\cite{fotoblok}.  
More generally, if the particle is related to a
super selection rule (SSR)~\cite{ssrdef}, a reference frame
(if such is available) can be used to locally break the SSR~\cite{reffra}, which
enables generation of the superposition.
 
\begin{figure}
\includegraphics[width = 0.45\textwidth]{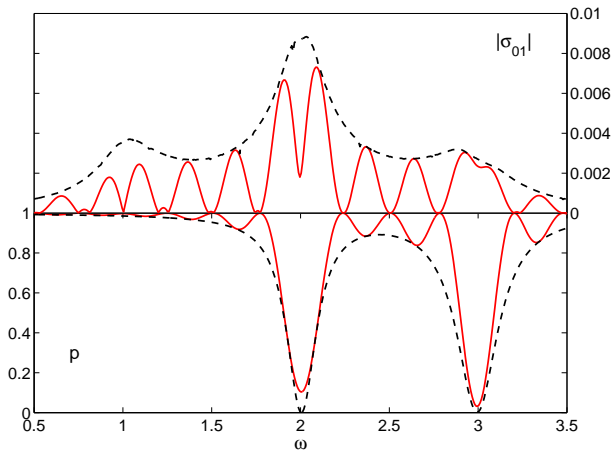}
\caption{\label{fig:treabs} (Color online) A photon interacting with a
  three-level atom, according to the Hamiltonian in Eq.~(\ref{trenivaa}). 
  The solid line in the lower panel shows the
  probability $p$ to detect the photon at time $t=25$ as a function of the photon
  energy $\omega$ measured in units of a reference energy $\mathcal{E}$. 
  The solid line in the upper panel shows the amount of superposition  
  $|\sigma_{01}|$ between the vacuum and a single photon at 
  $t=25$ (in units of $\mathcal{E}^{-1}$) as a
  function of $\omega$. 
  The dashed line in the upper (lower) panel is obtained by at each $\omega$ take the maximum 
  (minimum)  of $|\sigma_{01}|$  ($p$) over a long evolution time.}
\end{figure}

The following Hamiltonian of a photon and three level atom interaction 
results in a nontrivial superposition creation spectrum:
\begin{eqnarray}
\label{trenivaa}
H & = & \frac{\omega}{2}\sigma_{z}\otimes \hat{1}_{a} +
\hat{1}\otimes(\omega_{0}|0\rangle\langle0| + \omega_{1}|1\rangle\langle 1|
+ \omega_{2}|2\rangle\langle 2| )\nonumber\\
&&+ \sum_{k,k':k>k'}g_{k,k'}(\sigma_{-}\otimes|k\rangle\langle k'|
+ \sigma_{+}\otimes|k'\rangle\langle k|),
\end{eqnarray}
where $\sigma_{-} = |0\rangle\langle 1|$ and $\sigma_{+} = |1\rangle\langle
0|$, and where $\omega$, $\omega_{k}$, and $g_{k,k'}$ are in units of $\mathcal{E}$. 
Initially we have a single photon and the atom in its ground
state. To illustrate the effect we choose 
$\omega_{0} = 5$, $\omega_{1} = 7$, $\omega_{2} = 8$, $g_{01} =
0.05$, $g_{02} = 0.07$, and $g_{12} = 0.08$.   
In Fig.~\ref{fig:treabs} the solid lines give the probability $p$ to find a photon (lower panel) and
 the superposition creation $|\sigma_{01}|$ (upper panel) as functions of $\omega$ at 
$t=25$. 
In the upper panel the dashed line is the maximum of $|\sigma_{01}|$ at each $\omega$, 
taken over a long evolution time, thus approximating the envelope of the evolution 
of the $|\sigma_{01}|$. In the lower panel the dashed line similarly depicts 
the minimum of $p$.
As expected, there are two lines in the absorption spectrum corresponding to the two transitions
from the ground state. The superposition creation spectrum, however, 
shows a peak also at the transition between the two excited states. 
Further investigations are needed to understand the significance of this type of spectrum.

In conclusion we consider the coherence of single particles under particle loss.
Channels with no gain are not only characterized by the 
loss they cause, but also by how well they preserve coherence, 
and their tendency to create superposition between vacuum and single particle states.
We find an inequality relating these quantities.
This characterization of loss processes suggests a generalized 
spectroscopic approach
 where we record a ``coherence loss spectrum" 
and a ``superposition creation spectrum", akin to absorption spectra. 
We illustrate these concepts with examples.
Although the vacuum  preservation condition simplifies the analysis, 
the notions of coherence loss spectra and
superposition creation spectra are not limited to this setting. However, the
general case requires a more extensive analysis, and will most likely lead to
richer phenomena. A comparison with coherence spectroscopy in dissipative media  
\cite{cohspe, cohspediss}, and coherent control techniques in 
general \cite{cohcontr}, may also be fruitful.

J{\AA}. wishes to thank the Swedish Research Council for financial support
and the Centre for Quantum Computation at DAMTP, Cambridge, for hospitality.
DKLO acknowledge support from the Scottish Universities Physics Alliance.
This work was supported by the European Union through the Integrated Project
QAP (IST-3-015848), SCALA (CT-015714), SECOQC and the QIP IRC (GR/S821176/01).

\end{document}